\documentclass[aps,prb,reprint,a4paper,floatfix,superscriptaddress,amsmath,amssymb,amsfonts]{revtex4-1}
\usepackage{newtxtext,newtxmath}
\usepackage[utf8]{inputenx} 
\usepackage{graphicx}
\usepackage{tabularx, booktabs}
\usepackage{booktabs}
\usepackage{xspace}
\usepackage[dvipsnames]{xcolor}
\usepackage[pdftex]{hyperref}
\usepackage{color}
\usepackage[T1]{fontenc}
\hypersetup{
	pdftitle={Unravelling the oxygen influence in cubic bixbyite In$_2$O$_3$ on Raman active phonon modes by isotope studies},
	colorlinks,
	citecolor=blue
	}
\usepackage[
,textwidth=17.5cm
,textheight=23.7cm
,verbose
,dvips
]{geometry} 

\newcommand{\comment}[1]{}

\begin{document}

\title{Unravelling the oxygen influence in cubic bixbyite In$_2$O$_3$ on Raman active phonon modes by isotope studies}

\author{Johannes Feldl}
\affiliation{Paul-Drude-Institut f\"ur Festk\"orperelektronik, Leibniz-Institut im Forschungsverbund Berlin e.\,V., Hausvogteiplatz 5--7, 10117 Berlin, Germany}
\author{Roland Gillen}
\affiliation{Chair of Experimental Physics, Friedrich-Alexander Universit\"at Erlangen-N\"urnberg, Staudtstr. 7, 91058 Erlangen, Germany}
\affiliation{College of Engineering, Swansea University, Swansea SA1 8EN, United Kingdom }
\author{Janina Maultzsch}
\affiliation{Chair of Experimental Physics, Friedrich-Alexander Universit\"at Erlangen-N\"urnberg, Staudtstr. 7, 91058 Erlangen, Germany}
\author{Alexandra Papadogianni}
\affiliation{Paul-Drude-Institut f\"ur Festk\"orperelektronik, Leibniz-Institut im Forschungsverbund Berlin e.\,V., Hausvogteiplatz 5--7, 10117 Berlin, Germany}
\author{Joe Kler}
\affiliation{Institute of Physical Chemistry, RWTH Aachen University, Aachen, Germany}
\author{Zbigniew Galazka}
\affiliation{Leibniz-Institut f\"ur Kristallz\"uchtung, Max-Born-Stra{\ss}e 2, 12489 Berlin, Germany}
\author{Oliver Bierwagen}
\email{bierwagen@pdi-berlin.de}
\affiliation{Paul-Drude-Institut f\"ur Festk\"orperelektronik, Leibniz-Institut im Forschungsverbund Berlin e.\,V., Hausvogteiplatz 5--7, 10117 Berlin, Germany}
\author{Manfred Ramsteiner}
\affiliation{Paul-Drude-Institut f\"ur Festk\"orperelektronik, Leibniz-Institut im Forschungsverbund Berlin e.\,V., Hausvogteiplatz 5--7, 10117 Berlin, Germany}

\begin{abstract}
In this study, we performed comprehensive investigations on the Raman active phonon modes in cubic bixbyite In$_2$O$_3$, an important oxide based, wide-bandgap semiconductor. Fundamental insights into the lattice dynamics are revealed, by determining the atomistic contribution to all modes and their frequencies by density functional perturbation theory calculations. Those simulations were performed for different compositions of $^{16}$O and $^{18}$O isotope ratios, including their pure states. An increasing red-shift of the mode frequencies with increasing $^{18}$O content for all modes, due to the increased atomic mass, is revealed. For the seven lowest energy modes, this relative shift is below 1\%, whereas for the remaining 15 higher energetic modes a shift of about 5.5\% was identified. All modes have energy contributions of both indium and oxygen lattice sites, except for one, which corresponds to a pure oxygen vibrational state. Applying Raman spectroscopy, those results could be verified experimentally with excellent agreement. Investigated samples consisted of a bulk single crystal with $^{16}$O isotopes and a MBE grown thin film as the $^{18}$O sample. Time-of-flight secondary ion mass spectrometry confirms the purity of the oxygen isotope in the sample. These isotopologue studies allow for a direct experimental access to fundamental material properties in cubic In$_2$O$_3$ by means of Raman spectroscopy. For example, we speculate, that the presence of oxygen vacancies in In$_2$O$_3$ would result in a shift of modes that are dominated by O-vibrations, \textit{e.~g.}, $E_{g}^{(4)}$ or $A_{g}^{(4)}$, towards lower frequencies.

\end{abstract}

\maketitle

\section{\label{sec:level1}Introduction\protect\\}

\begin{figure}[h]
\includegraphics*[width=0.44\textwidth]{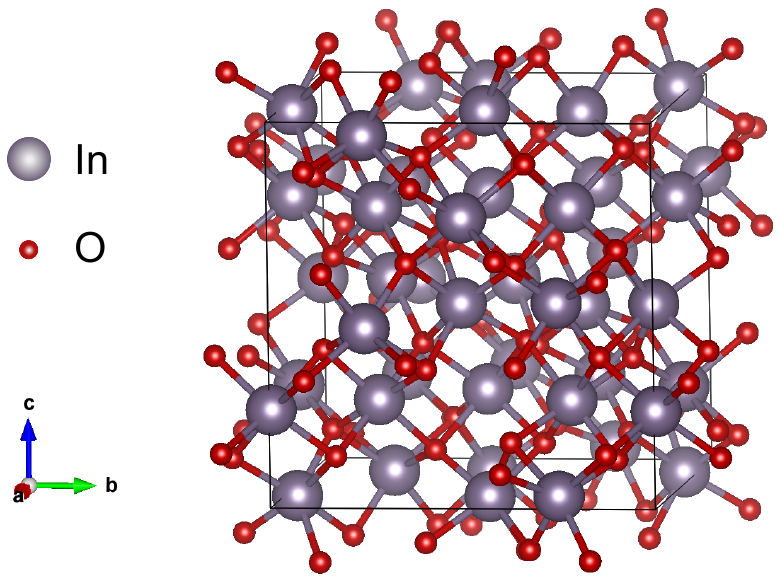}
\caption{Unit cell of cubic bixbyite In$_{2}$O$_{3}$, having space group $Ia\bar{3}$ (206). It comprises of 48 oxygen and 32 indium atoms, a lattice parameters of $a = b = c = 10.117$~\r{A} and $\alpha = \beta = \gamma = 90$°. The image was created using VESTA.\citep{momma2011}} 
\label{crystalstruc}
\end{figure}

The sesquioxide In$_2$O$_3$ is a wide-bandgap material with a fundamental bandgap in the range of 2.7 to 2.9~eV,\citep{janowitz2011, scherer2012, walsh2008, king2009} but a strong onset of optical absorption only occurring at 3.8~eV,\citep{feneberg2016} making it a promising material for transparent electronics.\citep{bierwagen2015, galazka2018, vonwenckstern2017} In$_2$O$_3$ is observed in two different polymorphes, a stable body centered cubic bixbyite and a meta-stable rhombohedral structure. Latter one is only observed when grown under special conditions\citep{wang2008} and will be disregarded for our discussion here. The bixbyite structure has space group 206 ($Ia\bar{3}$), a lattice parameter of $a = 10.117$~\r{A} and a cubic unit cell containing 16 formula units,\citep{marezio1966} making a total of 80 atoms, illustrated in Figure~\ref{crystalstruc}.
Widespread applications of In$_2$O$_3$ include heavily Sn doped (ITO) crystals with $n$-type conductivity \citep{chae2001, tiwari2004, tsai2016} and conductometric gas sensors.\citep{barsan2001b} For both applications, low quality crystals suffice. However, low crystal quality in In$_2$O$_3$ \textit{e.~g.} comes with a considerable high unintentially doped (UID) $n$-type conductivity or surface electron accumulation layer (SEAL), directly impacting the gas sensing properties of In$_2$O$_3$ and restricting the material's usecases.\citep{barsan2001b, dewit1977}  In recent years, sesquioxides attracted focused efforts towards more advanced semiconductor applications, such as power electronics\citep{higashiwaki2014} or heterostructures, stacking several different oxides.\citep{yu2016}
Photon detection in the ultraviolet regime would benefit greatly by a tailored absorption edge in the active material. A combination of In$_2$O$_3$ with Ga$_2$O$_3$ in the form of an (In,Ga)$_2$O$_3$ alloy\citep{wenckstern2015, feldl2021, hill1974, zhang2014, swallow2021, maccioni2016} promises such a bandgap engineering up to the 4.8~eV\citep{ueda1997, yamaga2011} of pristine Ga$_2$O$_3$. Along with such applications, semiconductor-grade crystal qualities are demanded. Inherently, a detailed understanding of the underlying fundamental mechanics of In$_2$O$_3$ is required and needs more studies to facilitate the rise in quality standards. A comprehensive understanding of phonon modes in In$_2$O$_3$ is integral to key material properties including mechanical strain, disorder, thermal and electrical dynamics as well as phonon-mediated optical processes. Hence, insights into vibrational properties facilitate improved crystal synthesis and offer critical input for both fundamental material characterization and applied studies, driving advances in both theoretical modelling and practical implementations of In$_2$O$_3$-based technologies.

The aim of the present work is to provide valuable insights into the fundamental mechanics of the lattice vibrations of cubic bixbyite In$_2$O$_3$ and the contribution of individual lattice sites to those. Raman spectroscopy is a powerful, experimental method for the investigation of the vibrational properties of In$_2$O$_3$.\citep{carvalho2018, regoutz2015, garcia-domene2012, feldl2021, kranert2014a} Aside from experimental approaches, theoretical methods, \textit{e.~g.} based on density functional theory frameworks, have been successfully applied to predict the Raman active phonon mode frequencies in In$_2$O$_3$\citep{garcia-domene2012} or for calculations on the fundamental bandgap of In$_2$O$_3$.\citep{peelaers2015} In the case of In$_2$O$_3$, vibrational modes could originate solely from indium or oxygen lattice sites, or a combination of those. Discrimination of a mode against these types of origin poses challenges, hence making an identification of the role of oxygen atoms in the phonon mode spectrum challenging.
 In our study, we combine confocal Raman spectroscopy experiments with complementary density functional perturbation theory (DFPT) studies, both with the novel approach for In$_2$O$_3$ of using different oxygen isotopes. As Raman vibrations are quantized oscillations between atoms of the crystal lattice, exchanging an $^{16}$O with its $^{18}$O isotope does not affect the electronic structure and preserves the crystal structure and phonon modes, but changes the frequencies of phonon modes, involving an oxygen lattice site, due to the change in mass. Given the relative low mass of oxygen atoms, the $^{18}$O isotope has a significant relative mass increase of 12.5\% over the $^{16}$O isotope. Therefore, using different oxygen isotopologues renders an excellent option to study isotope effects on phonon modes via theoretical calculations and Raman spectroscopy. This has been demonstrated for TiO$_2$,\citep{kavan2011, frank2012} where modes without oxygen contribution could be identified. A recent study by Janzen $\textit{et al.}$ (see Ref.~\onlinecite{janzen2021}) successfully applies DFPT calculations and Raman spectroscopy on $\beta$-Ga$_2$O$_3$ with $^{16}$O and $^{18}$O isotopologues, assigning the energy contribution of each lattice site to the vibrational modes. By extending this method to In$_2$O$_3$, our study aims on gaining valuable insights into the lattice dynamics of bixbyite In$_2$O$_3$ beyond the symmetry and frequencies of the Raman active phonon modes,\citep{kranert2014a, garcia-domene2012} by providing quantitative insights into the element specific energy contribution to those lattice vibrations.

\begin{figure}[h]
\includegraphics*[width=1.00\columnwidth]{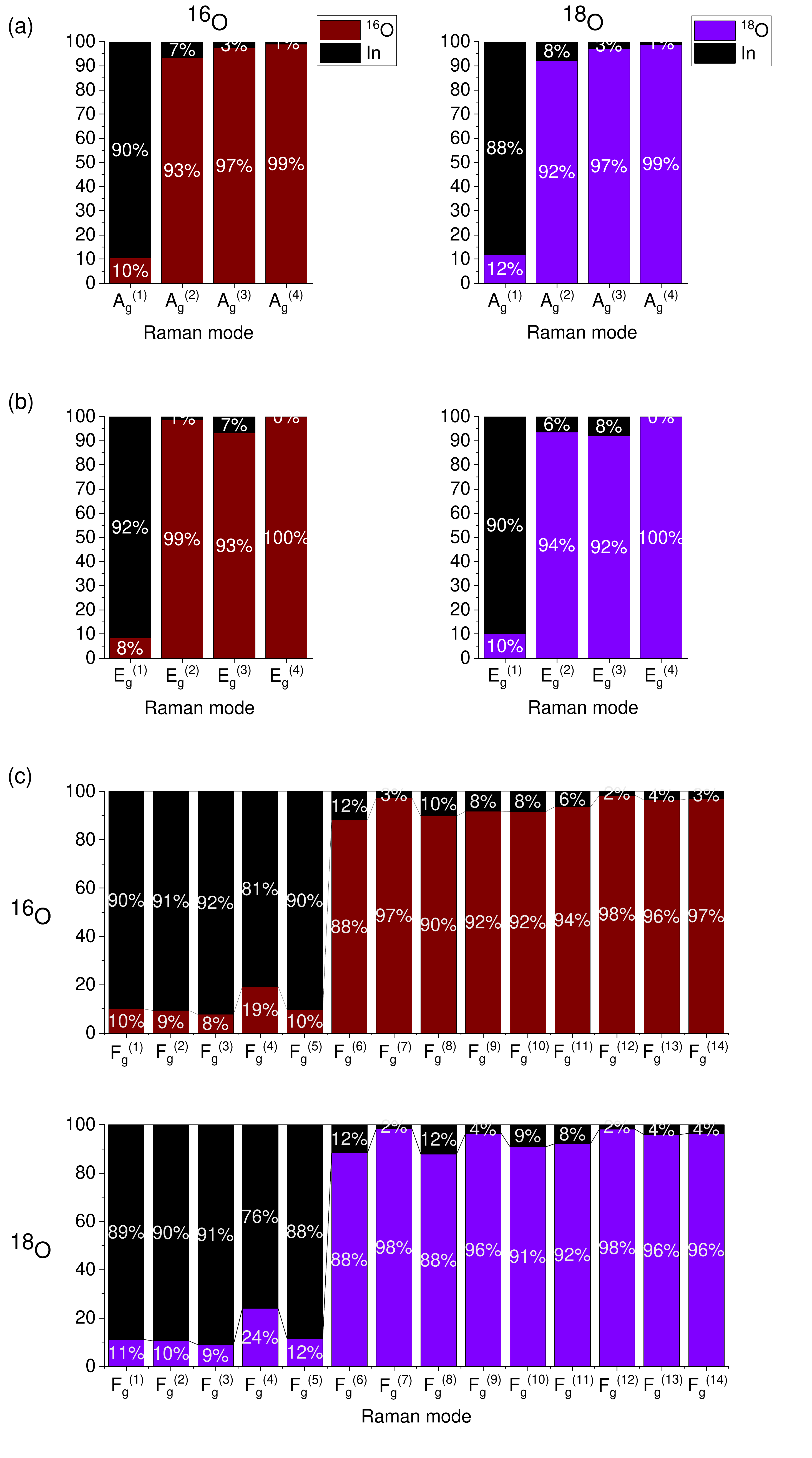}
\caption{Theoretical predictions for the relative energy contributions of indium (black) or oxygen, distinguished by isotope variants $^{16}$O (red) and $^{18}$O (purple), to the Raman-active lattice vibrations are presented. The analysed modes exhibit are modes with \textit{(a)} $A_{g}$, \textit{(b)} $E_{g}$ and \textit{(c)} $F_{g}$ symmetry. The relative energy contribution of each element is represented by the height of the corresponding coloured column. All modes consist of vibrations involving both elements present in the crystal, with the exception of mode $E_{g}^{(4)}$, which is purely oxygen-driven.} 
\label{AgEgFgmodes}
\end{figure}

\section{EXPERIMENTAL METHODS}

\begin{table*}
\caption{\label{PhononFrequencies}Experimentally and theoretically determined phonon mode frequencies of all Raman-active phonons in cubic bixbyite In$_{2}$O$_{3}$. The table lists frequencies for different oxygen isotope compositions In$_{2}$ $^{16}$O$_{3x}$ $^{18}$O$_{3(1-x)}$ as a function of $x$. Experimentally, two samples with a pure isotope composition of either 100\% $^{16}$O or $^{18}$O were investigated. Theoretical modeling additionally considered three mixed isotope configurations. The right section of the table presents the absolute and relative Raman mode frequency shifts $\Delta\omega$ for the fully substituted $^{16}$O and $^{18}$O samples, based on both experimental and theoretical wavenumbers.}
\begin{ruledtabular}
\begin{tabular}{c c c c c c c c c | c c c c}
 {} & {} & \multicolumn{2}{c|}{Experiment} & \multicolumn{5}{c|}{Theory}  & \multicolumn{2}{c}{Experiment} & \multicolumn{2}{c}{Theory} \\
{} & $x$ & 1 & \multicolumn{1}{c|}{0} & 1 & 0.75 & 0.5 & 0.25 & 0 & \multicolumn{4}{c}{$\Delta\omega$} \\
\hline
Mode & {} & (cm$^{-1}$) & (cm$^{-1}$) & (cm$^{-1}$) & (cm$^{-1}$) & (cm$^{-1}$) & (cm$^{-1}$) & (cm$^{-1}$) & (cm$^{-1}$) & (\%) & (cm$^{-1}$) & (\%) \\
\hline
 $F_{g}^{(1)}$ & {} & 112.3 & 111.6 & 106.5 & 106.4 & 106.3 & 106.2 & 106.1 & 0.7 & 0.62 & 0.4 & 0.38 \\
 $F_{g}^{(2)}$ & {} & 121.5 & 120.5 & 115.3 & 115.2 & 115.1 & 114.9 & 114.8 & 1.0 & 0.82 & 0.5 & 0.43 \\
 $A_{g}^{(1)}$ & {} & 135.3 & 134.1 & 129.9 & 129.7 & 129.5 & 129.3 & 129.1 & 1.2 & 0.89 & 0.8 & 0.62 \\
 $F_{g}^{(3)}$ & {} & 156.5 & 155.5 & 150.6 & 150.5 & 150.3 & 150.1 & 150.0 & 1.0 & 0.64 & 0.6 & 0.40 \\
 $E_{g}^{(1)}$ & {} & 172.7 & 171.7 & 167.4 & 167.2 & 166.9 & 166.7 & 166.5 & 1.0 & 0.58 & 0.9 & 0.54 \\
 $F_{g}^{(4)}$ & {} & 209.8 & 208.6 & 203.3 & 203.0 & 202.7 & 202.4 & 202.0 & 1.2 & 0.57 & 1.3 & 0.64 \\
 $F_{g}^{(5)}$ & {} & 215.6 & 214.4 & 208.2 & 208.1 & 207.9 & 207.7 & 207.5 & 1.2 & 0.56 & 0.7 & 0.34 \\
 $F_{g}^{(6)}$ & {} & 303.4 & 287.1 & 305.1 & 300.7 & 296.4 & 292.3 & 288.4 & 16.3 & 5.37 & 16.7 & 5.47 \\
 $A_{g}^{(2)}$ & {} & 309.4 & 293.1 & 306.8 & 302.4 & 298.2 & 294.1 & 290.3 & 16.3 & 5.27 & 16.5 & 5.38 \\
 $E_{g}^{(2)}$ & {} & 313.3 & 297.0 & 313.4 & 308.7 & 304.2 & 299.8 & 295.7 & 16.3 & 5.20 & 17.7 & 5.65  \\
 $F_{g}^{(7)}$ & {} & 318.0 & 302.2 & 316.6 & 311.8 & 307.3 & 303.0 & 298.8 & 15.8 & 4.97 & 17.8 & 5.62 \\
 $F_{g}^{(8)}$ & {} & 369.2 & 349.3 & 362.8 & 357.5 & 352.6 & 347.9 & 343.3 & 19.9 & 5.39 & 19.5 & 5.37 \\
 $F_{g}^{(9)}$ & {} & 392.1 & 371.2 & 385.1 & 379.6 & 374.3 & 369.3 & 364.4 & 20.9 & 5.33 & 20.7 & 5.38 \\
 $E_{g}^{(3)}$ & {} & 396.0 & 374.2 & 390.9 & 385.3 & 380.0 & 374.9 & 370.1 & 21.8 & 5.51 & 20.8 & 5.32 \\
 $F_{g}^{(10)}$ & {} & 455.1 & 431.2 & 442.0 & 435.5 & 429.3 & 423.4 & 417.8 & 23.9 & 5.25 & 24.2 & 5.48 \\
 $F_{g}^{(11)}$ & {} & 468.9 & 444.8 & 454.7 & 448.2 & 442.0 & 436.1 & 430.5 & 24.1 & 5.14 & 24.2 & 5.32 \\
 $A_{g}^{(3)}$ & {} & 497.2 & 472.6 & 481.7 & 474.5 & 467.7 & 461.2 & 455.0 & 24.6 & 4.95 & 26.7 & 5.54 \\
 $F_{g}^{(12)}$ & {} & 518.0 & - & 506.4 & 498.7 & 491.5 & 484.5 & 477.9 & - & - & 28.5 & 5.63 \\
 $F_{g}^{(13)}$ & {} & 546.9 & - & 529.4 & 521.6 & 514.1 & 506.8 & 500.0 & - & - & 29.4 & 5.63 \\
 $E_{g}^{(4)}$ & {} & 595.3 & 562.7 & 587.4 & 578.6 & 570.1 & 562.0 & 554.3 & 32.6 & 5.48 & 32.9 & 5.71 \\
 $A_{g}^{(4)}$ & {} & 608.9 & 575.5 & 575.9 & 567.1 & 558.7 & 550.7 & 543.0 & 33.4 & 5.49 & 33.1 & 5.64 \\
 $F_{g}^{(14)}$ & {} & 631.8 & 597.7 & 611.9 & 602.8 & 594.2 & 585.9 & 578.0 & 34.1 & 5.40 & 33.9 & 5.54 \\

\end{tabular}
\end{ruledtabular}
\end{table*}

The investigated 0.5~$\mu$m thin In$_2^{18}$O$_3$ film 
was grown homoepitaxially on the (1~1~1) surface of an In$_2$O$_3$ substrate by plasma-assisted molecular beam epitaxy (PA-MBE) at a growth temperature of 750~°C, with the oxygen-plasma source running at 300~W. As oxygen source, a nominally 97.39\% enriched $^{18}$O$_2$ isotope source was employed, with an oxygen flux of 0.3~sccm  during the film deposition. For the homoepitaxial growth, a high quality single crystal bulk In$_2$O$_3$, with a thickness of 500~$\mu$m, was used. In$_2$O$_3$ substrates for homoepitaxy were prepared from bulk single crystals grown directly from the melt at the Leibniz-Institut f\"ur Kristallz\"uchtung, Berlin, using a novel crystal growth technique under the name \textit{Levitation-Assisted Self-Seeding Crystal Growth Method}.\citep{galazka2014}

Structural characterization of the deposited film succeeded by XRD measurements (PANalytical X’Pert Pro MRD equipped with a four-axis goniometer and using well-collimated and monochromatic Cu K$\alpha_1$ radiation). Symmetric, out-of-plane $2\Theta-\omega$-scans and $\omega$-rocking curves of the In$_2$O$_3$ 2~2~2 reflex were performed to investigate the out-of-plane lattice parameter as well as the strain relaxation of the layer. The out-of-plane lattice parameters were additionally measured by high-resolution XRD (HRXRD).

For the determination of the actual incorporated $^{18}$O isotope ratio into the film's In$_2$O$_3$ matrix, time-of-flight secondary ion mass spectrometry (ToF-SIMS) was performed on a ToF-SIMS IV machine (IONTOF GmbH, Münster, Germany) equipped with a ToF-SIMS V analyser and an extended dynamic range (EDR) unit. A high energy Ga$^+$ primary ion gun, operated at 25~keV and in spectrometry mode, was rastered on an area of 100~µm~$\times$~100~µm to produce secondary ions for ToF analysis. Negative secondary ions were collected with a cycle time of 50~µs. The EDR was used for the $^{16}$O$^-$ and $^{18}$O$^-$ species to prevent overloading the detector. A low energy Cs$^+$ ion gun, operated at 2~keV, was used to sputter etch the sample over an area of 400~µm~$\times$~400~µm. Charge compensation was achieved by using a low energy ($<20$~eV) electron flood gun. The crater depths were measured post analysis, using an Interference Microscope (NT1100, Veeco Instruments Inc., NY, USA).

Raman spectroscopic measurements were performed in backscattering geometry with optical excitations at the 325-nm (3.81~eV) line of a He-Cd laser. For polarization-dependent mode symmetry identification, a second setup with an excitation wavelength of 473~nm (2.62~eV) of a solid-state laser was used. In both cases, the incident laser light was focused by a microscope objective onto the sample surface. The backscattered light was collected by the same objective, spectrally dispersed by a 80-cm spectrograph (LabRam HR, Horiba/Jobin Yvon) and detected by a liquid-nitrogen-cooled charge-coupled device (CCD). Calibrating the Raman spectra was carried out by a reference Hg gas-discharge lamp (3.81~eV spectra) and a silicon reference sample (2.62~eV spectra). All spectra were acquired at room-temperature. Polarization dependent Raman spectra were recorded with the utilization of $\lambda/2$ waveplates and polarization analysers. Using the Porto notation,\citep{damen1966} the $z(xx)\bar{z}$ and $z(xy)\bar{z}$ scattering geometries were applied, where $z$ denotes the direction normal to the sample surface and $x$, $y$ spanning the surface plane, perpendicular to each other. For cubic bixbyite In$_2$O$_3$ in [111] orientation, it is known, that a $z(xy)\bar{z}$ configuration leads to an extinction of modes with $A_g$ symmetry, while the $z(xx)\bar{z}$ setting reduces the $F_g$ modes intensity by a third.\citep{kranert2014a}
For the determination of phonon mode frequencies, the Raman spectra were analysed by line-fittings with Voigt functions resulting in an accuracy of 0.1~cm$^{-1}$ for the peak positions.

\section{THEORETICAL METHODS}

Simulations of the lattice vibrations were performed within the frame of density functional perturbation theory (DFPT) on the level of the Perdew-Burke-Ernzerhof exchange-correlation approximation for solids (PBEsol) as implemented into the Quantum Espresso computational package.~\citep{giannozzi2009} The In (5s,5p,4d) and the O (2s,2p) states were treated as valence electrons using multi-projector optimized normconserving Vanderbildt (ONCV) pseudopotentials~\citep{hamann2013} from the Pseudo Dojo repository~\citep{vansetten2018} and described by a planewave basis set using a cutoff energy of 180\,Ry (2450\,eV). All reciprocal space integrations were performed by a discrete $k$-point sampling of 6×6×6 points in the Brillouin zone. We fully optimized the atomic positions and cell parameters of the primitive cell cubic In$_2$O$_3$ the residual forces between atoms and the cell stress were smaller than 0.0025\,eV\r{A}$^{-1}$ and 0.01\,GPa, respectively. The threshold for the total energy was set to 10$^{-14}$\,Ry, which ensured tightly converged interatomic forces for the geometry optimization and of the ground state density and wavefunctions for the DFPT calculations. We obtain a lattice constant of $a=10.17$~\r{A}, \textit{i.e.} an underbinding of 0.5\% compared to the experimental lattice constant ($a^{\mathrm{exp}}=10.119$~\r{A}) and a corresponding systematic underestimation of the simulated Raman shifts. Using the obtained phonon mode frequencies and displacement patterns, we then computed the relative energy contributions of indium and oxygen atoms to the Raman active lattice vibrations for In$_2$O$_3$ systems containing $^{16}$O and $^{18}$O isotopes using the method outlined in Ref.~\onlinecite{janzen2021}.

\section{RESULTS and DISCUSSION}
\subsection{\label{sec:leve31}{Density functional perturbation theory}}

Cubic bixbyite In$_2$O$_3$ contains 24 oxygen and 16 indium atoms per primitive unit cell. Group theory predicts for the given space group 206 ($Ia\bar{3}$) a total of 120 vibrational modes.\citep{garcia-domene2012, white1972} Regarding only the \textit{gerade} (\textit{g}) modes, which are Raman active, a total of 22 modes remain, which can be represented at the $\Gamma$-point as following:

\begin{equation}
\Gamma_{g} = 4A_{g} + 4E_{g} + 14F_{g},
\label{gerade_modes}
\end{equation} 

where $E_{g}$ modes are double and $F_{g}$ modes triple degenerated.\citep{stokey2021}

The resulting energy contribution of In and O atoms to $A_{g}$, $E_{g}$ and $F_{g}$ modes can be seen in Figure~\ref{AgEgFgmodes}~(a)-(c). 
A notable finding is that, for all mode symmetries, there is at least one low-energy mode with approximately 10\% of the total mode energy stemming from the oxygen lattice sites. Conversely, several high-energy modes exhibit over 90\% of their vibrational energy originating from oxygen related vibrations. Most strikingly, when plotted as a function of mode frequency, these two distinct contribution regimes form two plateaus, characterized by a step-like behaviour, with no gradual redistribution occurring for modes with intermediate energy. Mode $E_{g}^{(4)}$ stands out compared to the rest of the 21 phonon active modes, since it is the only mode, where only one element (oxygen) is contributing to the vibration, see Figure~\ref{AgEgFgmodes}~(b). All other modes are a combination of In$-$O vibrations. A large oxygen related energy contribution translates into a strong sensitivity in terms of mode frequency and intensity with respect to changes in physical properties, correlated with oxygen lattice sites. No significant change in element specific energy contribution to the modes can be found for a crystal with $^{18}$O instead of $^{16}$O. In the same approach, a prediction of the absolute wavenumber of the modes were calculated for pure $^{18}$O and $^{16}$O crystals as well as of mixed isotope compositions (incremental steps of 25\% of relative isotope contents). Table~\ref{PhononFrequencies} lists the absolute wavenumbers of all phonon modes for both experimental and theoretical results, along with the absolute and relative frequency shifts ($\Delta\omega$) between the two isotope variants. The relative shifts are referenced to $^{16}$O. As one of the most relevant results of our calculations, a clear monotonic shift towards lower mode frequencies with increasing $^{18}$O content is revealed.

\subsection{\label{sec:leve32}{ToF-SIMS and XRD}}

An essential experiment for the entire discussion is the determination of the successful incorporation of the $^{18}$O isotope into the homoepitaxially grown In$_2$O$_3$ film. From our Raman experiments independent verification and quantification comes from ToF-SIMS experiments. The quantity of interest is the $^{18}$O isotope fraction $n^{*}$, which is directly reflected by the isotope intensities

\begin{equation}
n^{*} = \frac{I_{\mathrm{^{18}O}}}{I_{\mathrm{^{16}O}} + I_{\mathrm{^{18}O}}}.
\label{nstar}
\end{equation}

Throughout the $^{18}$O layer, a constant $^{18}$O isotope fraction of 96.5\% could be determined. After approximately 0.5~$\mu$m, $n^{*}$ drops below 7\%. This steep drop represents the film-substrate interface. These results allow for the assumption of a homogeneous and almost pure $^{18}$O presence for the Raman spectroscopic probing depth at 3.81~eV. The corresponding ToF-SIMS scan can be found in Figure S1 in the supplementary material.

XRD analysis of the $^{18}$O thin film on In$_2$O$_3$(1~1~1) substrate (see Fig.~S1 and related discussion in the supplementary material) indicates the exclusive presence of (1~1~1) oriented In$_2$O$_3$ lattice planes with a cubic lattice parameter of $a = 10.119$~\r{A} (within 0.02~\% of that given in \onlinecite{marezio1966}). A narrow $\omega$-rocking curve (full widths at half maximum 0.014~deg) indicates excellent crystal quality. The out-of-plane lattice mismatch of the film to the substrate is estimated to be within $\pm0.035$~\%, translating into a negligible in-plane strain state for the thin film. Due to the negligible lattice strain and the overall high quality of the film, no impact of the growth conditions on the phonon mode structure and the Raman selection rules are expected for the $^{18}$O containing sample.\citep{garcia-domene2012, liu2008}

\subsection{\label{sec:leve33}{Raman spectra}}

\begin{figure}[h]
\includegraphics*[width=0.5\textwidth]{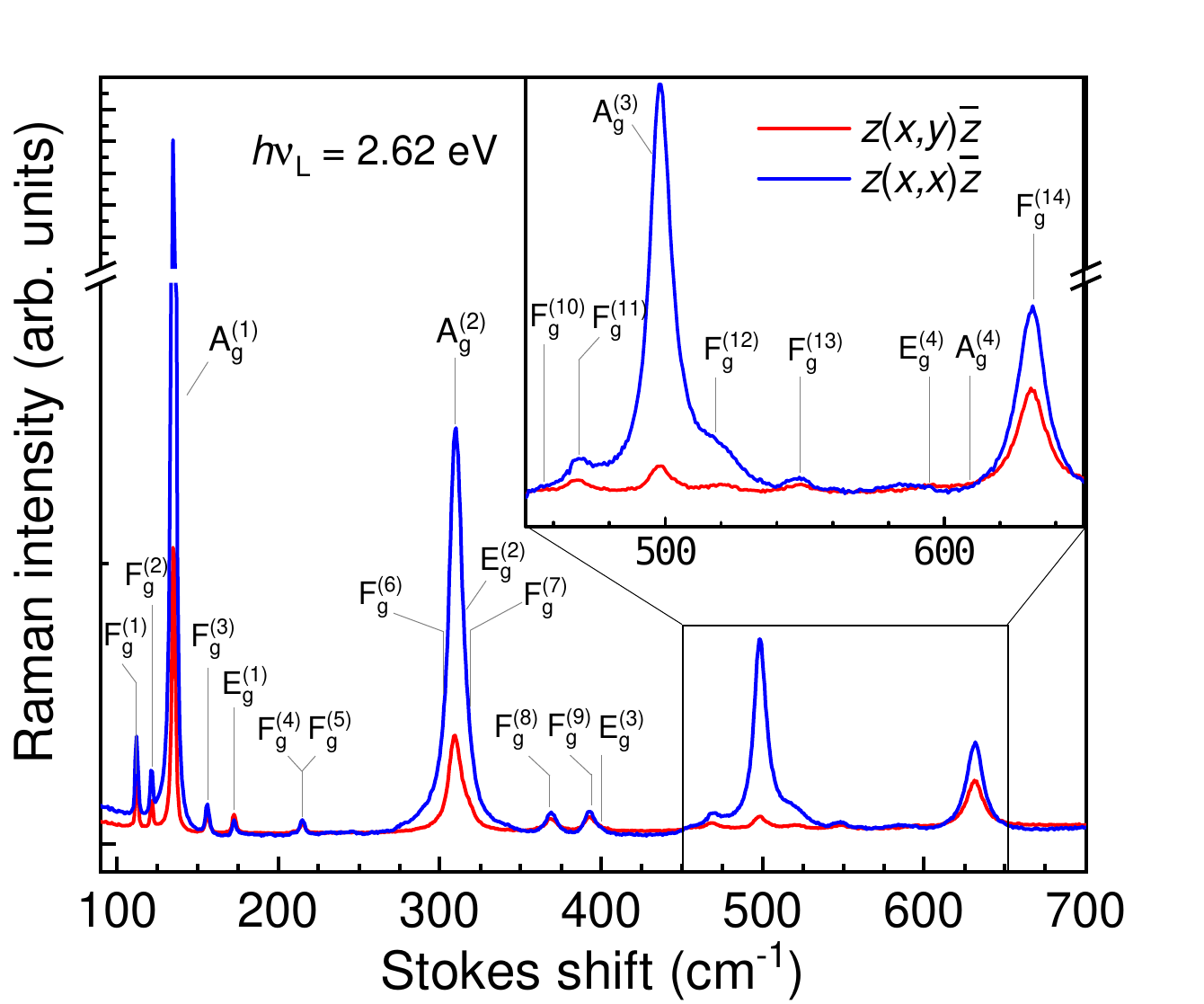}
\caption{Raman spectra of the $^{16}$O In$_{2}$O$_{3}$ sample, acquired at room temperature, 2.62~eV excitation photon energy and for two different polarization states, $z(xx)\bar{z}$ and $z(xy)\bar{z}$ following the Porto notation. As the $A_{g}$ modes intensity is highly dependent on the polarization state of the scattered Raman photon, this dependence can be applied for an unambiguous mode identification.} 
\label{PolarizationsPhonons}
\end{figure}

All theoretically predicted 22 Raman active phonon modes could be identified for the $^{16}$O bulk-like substrate, while for the $^{18}$O homoepitaxially grown film, a total of 20 Raman active modes were successfully identified. Thereby, we took advantage of the optical excitation at 2.62~eV, for which the optical probing depth $(\alpha_{i} + \alpha_{s})^{-1}$ in In$_2$O$_3$\citep{feneberg2016} is much larger than the film thickness, leading to an excellent signal-to-noise ratio. In particular, our bulk sample provides enough interaction volume and a resulting signal strength sufficient to detect the complete set of cubic bixbyite In$_2$O$_3$ phonon modes. Figure~\ref{PolarizationsPhonons} displays the corresponding Raman spectra of the $^{16}$O bulk sample measured with two different polarization configurations. The utilization of excitation in the transparent regime together with polarization-dependent measurements has been particularly beneficial for disentangling the Raman signal at around 310~cm$^{-1}$, a superposition of four Raman modes, $F_{g}^{(6)}$, $A_{g}^{(2)}$, $E_{g}^{(2)}$ and $F_{g}^{(7)}$. Assuming a negligible contribution of the $E_{g}^{(2)}$ mode for excitation at 2.62~eV,\citep{kranert2014a} a detailed line-fitting analysis of the Raman spectra measured with different scattering geometries allowed us to unambiguously attribute the central peak as well as a red and blue shifted shoulder peaks to the $F_{g}^{(6)}$ mode at 303.4~cm$^{-1}$, the $A_{g}^{(2)}$ at 309.4~cm$^{-1}$ and the $F_{g}^{(7)}$ one at 318.0~cm$^{-1}$, respectively. With the knowledge of these mode frequencies, the $E_{g}^{(2)}$ mode was determined with a wavenumber of 313.3~cm$^{-1}$ by a line-fitting analysis of Raman spectra excited at 3.81~eV, in excellent agreement with the numerical predictions and with unambiguously identifiable $^{16}$O modes in the study by Kranert \textit{et al.}\citep{kranert2014a}, cf. Table~\ref{PhononFrequencies}. Such a disentanglement of the superimposed Raman peaks at 310~cm$^{-1}$ has not been possible in previous reports based on the investigation of thin films\citep{kranert2014a} and is described in more detail in the supplementary material.
\\

In order to exclude any contribution from the substrate to the Raman signal when investigating the homoepitaxially grown $^{18}$O film, using an excitation energy above the bandgap is necessary. With an approximate film thickness of 0.5~$\mu$m and an excitation energy of 3.81~eV, which results in the small optical probing depth of 80~nm,\citep{feneberg2016} any contribution of the In$_2$O$_3$ substrate to the spectra, which would be present at an excitation at 2.62~eV, can be excluded.
Since our calculations predict a phonon mode specific, monotonic decrease of the phonon mode frequencies with increasing $^{18}$O content (see Table~\ref{PhononFrequencies}), the mode assignment for the $^{18}$O containing film could be derived in a straightforward manner from the spectrum of the $^{16}$O bulk sample.

\begin{figure}[h]
\includegraphics*[width=0.46\textwidth]{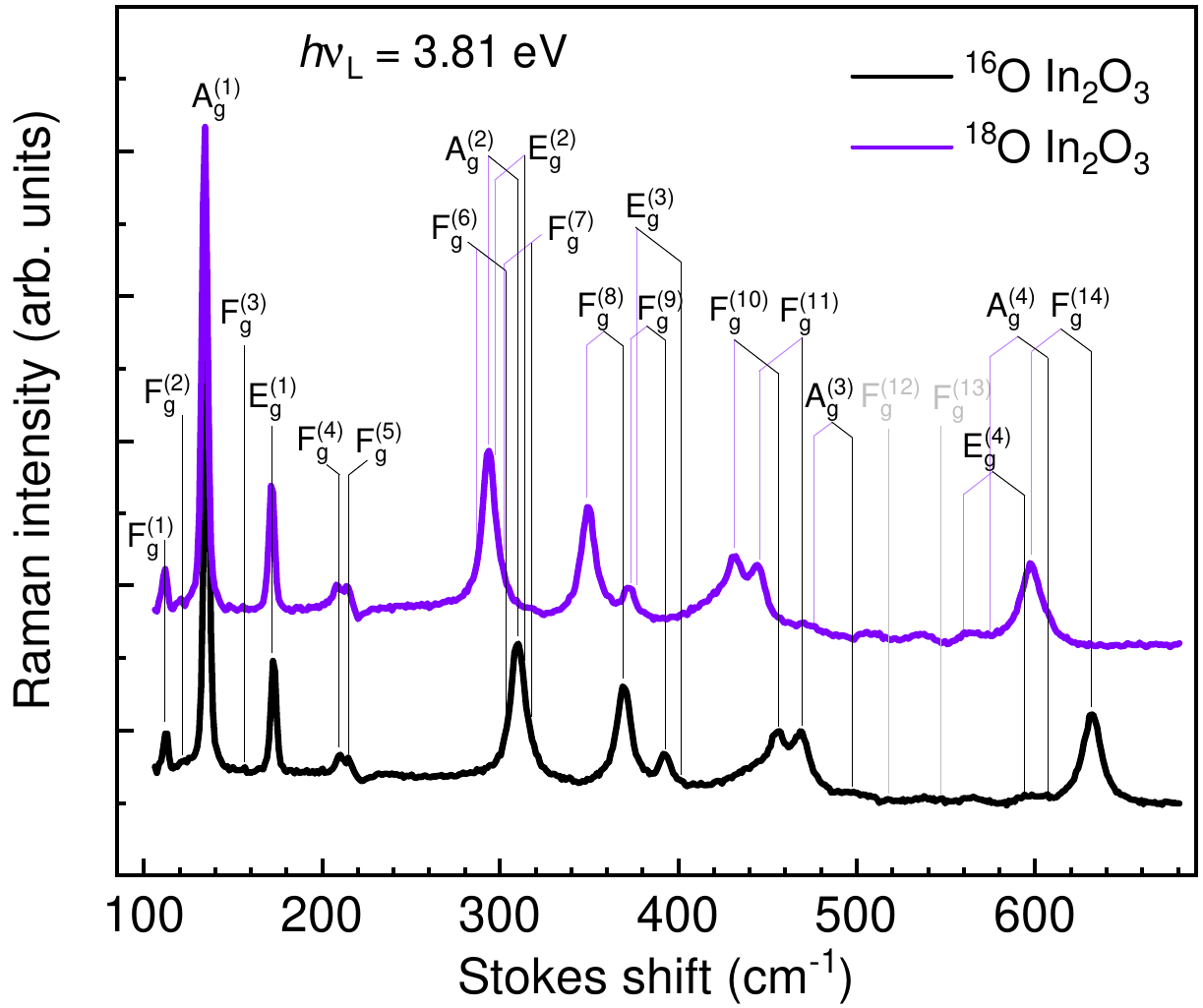}
\caption{Room temperature Raman spectra, acquired with an excitation photon energy of 3.81~eV, of the investigated cubic bixbyite In$_{2}$O$_{3}$ samples with natural, predominantly $^{16}$O and 96.5\% enriched $^{18}$O oxygen abundance, depicted in black and purple, respectively. The two spectra have an artificial offset, for a clearer optical distinguishably. Except for the $F_{g}^{(12)}$ and $F_{g}^{(13)}$ ones (light grey), all Raman active phonon modes could be identified for both spectra. A clear red-shift for the modes stemming from the heavier oxygen isotope is apparent for frequencies above 300~cm$^{-1}$.} 
\label{18Ovs16O} 
\end{figure}

\subsection{\label{sec:leve34}{Discussion of isotopic Raman shifts}}

\begin{figure}[h]
\includegraphics*[width=0.37\textwidth]{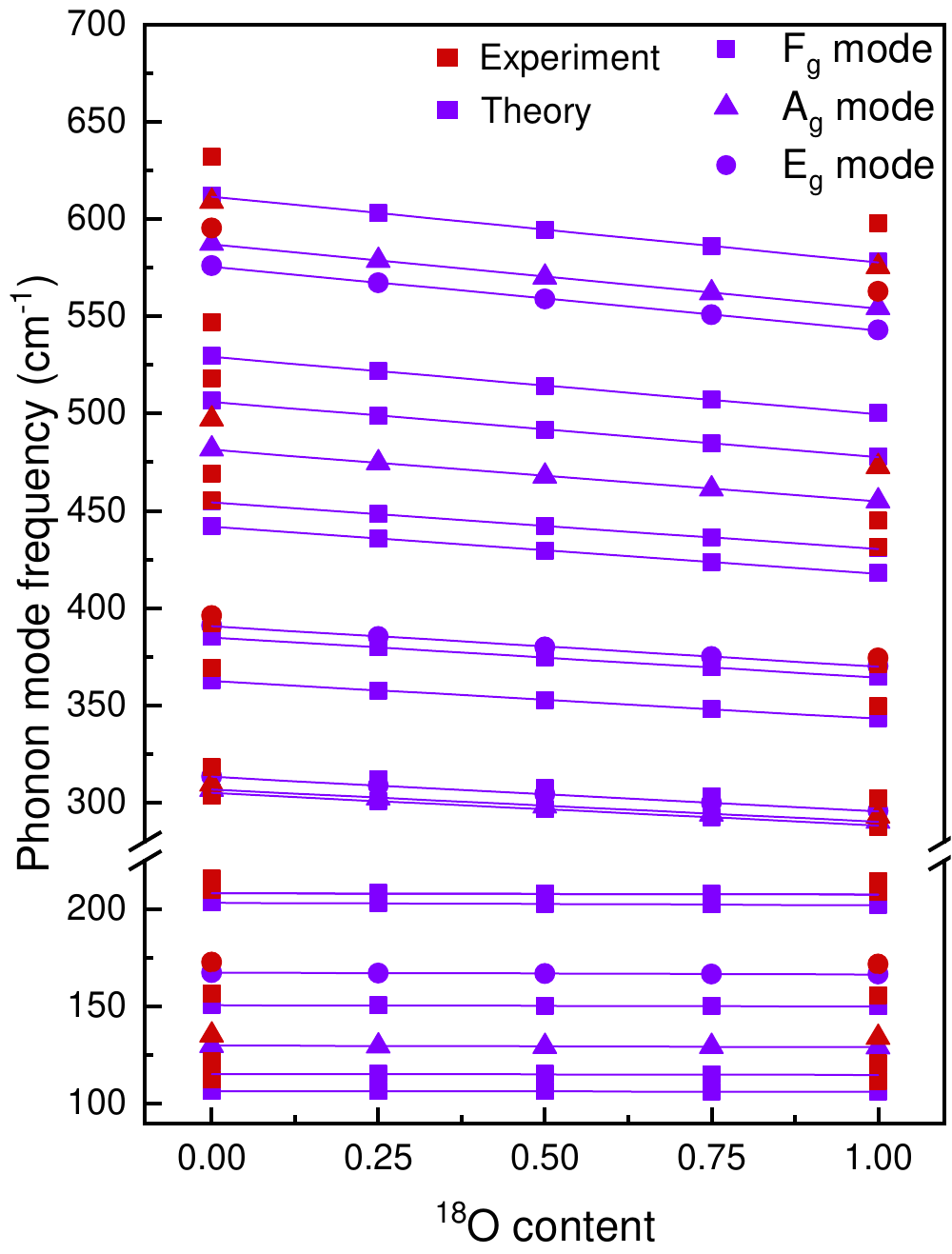}
\caption{Visualization of the Raman active phonon mode frequencies as a function of the $^{18}$O isotope content. $A_{g}$ modes are symbolized by rectangles, $E_{g}$ modes by circles and $F_{g}$ modes by squares. Compared are experimentally (red) determined wavenumbers with theoretically calculated (purple) values. The theoretical results underestimate the phonon mode wavenumbers slightly compared to the experimental results. However, both methods reveal the behaviour of a decrease in mode energy with rising $^{18}$O content for modes above 300~cm$^{-1}$ contrasting the almost constant behaviour below that threshold.} 
\label{IsotopeContent}
\end{figure}

The experimentally acquired Raman spectra, shown in Figure~\ref{18Ovs16O}, were recorded at room-temperature and with an excitation photon energy of 3.81~eV, whereby the probing volume is constrained within the MBE grown $^{18}$O containing film. Comparing the two In$_2$O$_3$ crystals with either $^{16}$O or $^{18}$O isotopologues, a striking effect of the increased oxygen mass is the red-shift of phonon modes with wavenumbers higher than approximately 300~cm$^{-1}$. Phonon modes with wavenumbers below 300~cm$^{-1}$ show a negligible dependence on the type of the oxygen isotope. Other aspects of the spectrum, such as the mode intensities and line-shape, do not differ with respect to the $^{16}$O grown crystal. Therefore, the experimental results support our theoretical predictions, indicating that oxygen vibrations contribute only a minor fraction - approximately 10\% - to the total mode energy of the first seven lowest-energy modes, see Figure~\ref{AgEgFgmodes}~\textit{(a)} -- \textit{(c)}.

The Raman mode frequencies of the epitaxial In$_2$O$_3$ film obtained by line-fitting are shown in Figure~\ref{IsotopeContent} as a function of the $^{18}$O content. Our prediction of a monotonic, linear decrease of the phonon mode frequencies with increasing $^{18}$O content could be confirmed for all resolved phonon modes, \textit{i.~e.} all but the $F_{g}^{(12)}$ and $F_{g}^{(13)}$ modes. Raman features observed in the corresponding range of 520 -- 580~cm$^{-1}$ for excitation energy of 3.81~eV are predominantly attributed to higher-order Raman scattering processes involving phonon modes with $A_{g}$ symmetry.\citep{kranert2014a}

The simple linear dependence of the mode frequencies on the isotope ratio could, for example, be utilized to study the fundamental MBE growth dynamics for bixbyite-type oxide systems when two different oxygen sources are used simultaneously.\citep{hoffmann2020} Supplying different oxygen isotopes by the two sources, their relative incorporation efficiencies can be determined by ex-situ Raman experiments by analyzing the phonon mode frequencies.

Table~\ref{PhononFrequencies} provides a detailed breakdown of the absolute and relative mode shifts for the two isotope variants based on experimental and theoretical results. Overall, the DFPT results underestimate the experimental mode frequencies by about 4\%, which is well within the typical accuracy of DFT simulations for phonons in oxides.\citep{he2014} The slight 'softening' of lattice vibrations is a known property of the PBEsol exchange-correlation functional used in our simulations and might indicate a somewhat overestimated screening of long-range interactions between atoms. For modes in the frequency range around 300~cm$^{-1}$ ($F_{g}^{(6)}$, $A_{g}^{(2)}$, $E_{g}^{(2)}$ and $F_{g}^{(7)}$), however, a reduced discrepancy between experimental and theoretical values is found. However, we note that the reliable experimental determination of the superimposed modes in this frequency range is challenging.

 The absolute frequency shifts, comparing both isotopologues, increase almost monotonically for both, the experimental and theoretical results, with increasing mode frequency. The range of experimentally observed shifts reaches from 0.7~cm$^{-1}$ to 34~cm$^{-1}$.
 
\begin{figure}[h]
\includegraphics*[width=0.42\textwidth]{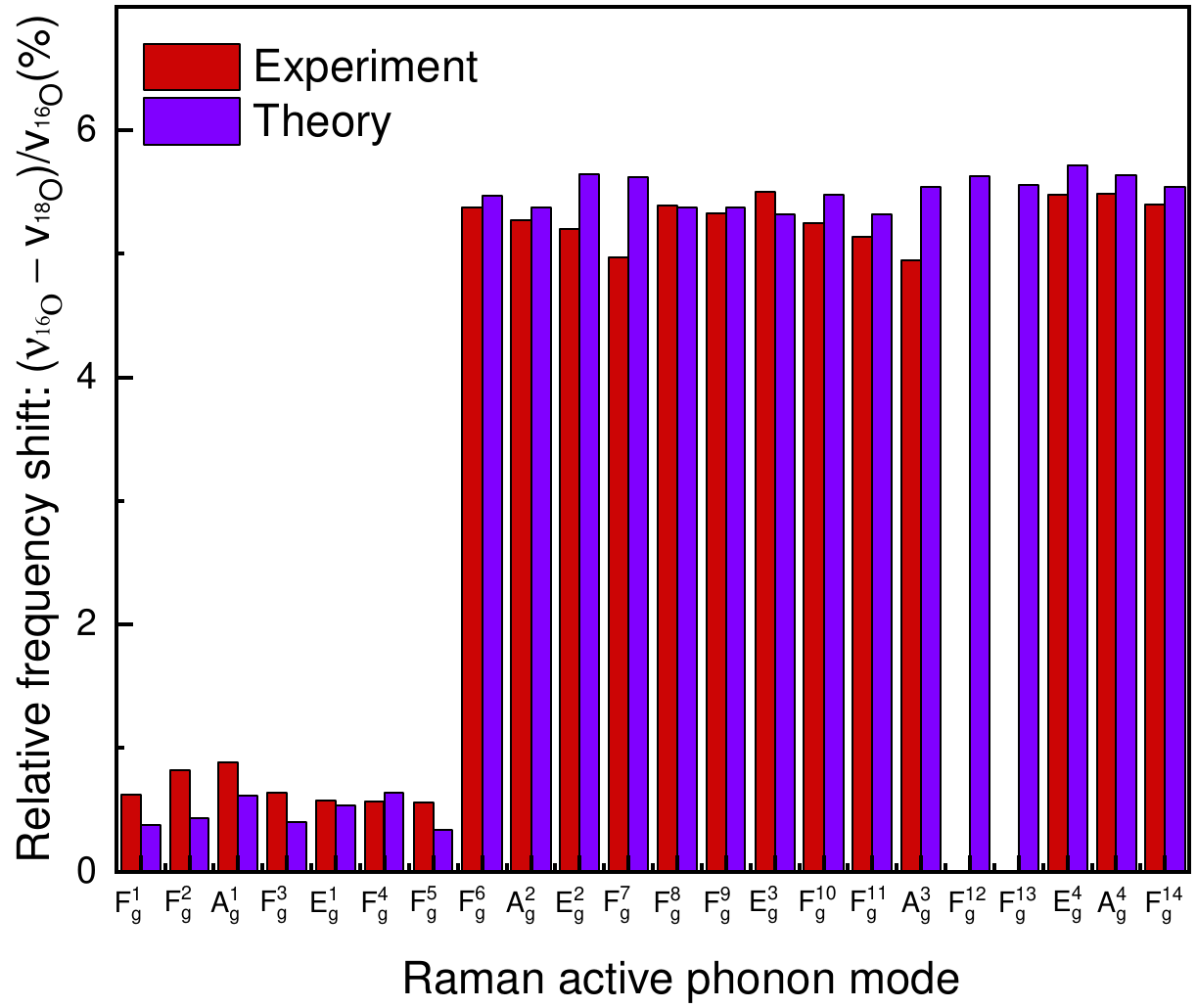}
\caption{Relative frequency shift of the Raman active phonon modes of the $^{16}$O to the $^{18}$O isotope sample, with the $^{16}$O values treated as the reference system. Both, experimental (red) and theoretical (purple) results are in great agreement with each other. Experimental values for modes $F_{g}^{(12)}$ and $F_{g}^{(13)}$ are missing, due to difficulties in their analysis. For modes with frequencies up to the value of the $F_{g}^{(5)}$ mode, show a very low oxygen isotope dependence of sub 1\% in relative frequency shift. In contrast, modes with higher frequencies show an almost constant relative shift of greater than 5\%.} 
\label{RelShift}
\end{figure}

The most striking observation, however, concerns the two plateaus in the relative frequency shifts $(^{16}\mathrm{O} - ^{18}\mathrm{O}) / ^{16}\mathrm{O})$. For modes with frequencies up to 215~cm$^{-1}$ ($F_{g}^{(5)}$), the relative frequency shifts remain on the order of 1\%, whereas for modes with higher frequencies, a values of 5\% and higher are found both experimentally and theoretically. Figure~\ref{RelShift} illustrates this behaviour including the excellent agreement between experiment and theory. Note, that no influence of the mode-symmetries on the relative frequency shifts could be identified. The two-plateau behaviour, however, corresponds exactly to the result of our calculations shown in Figure~\ref{AgEgFgmodes} and is, thus, explained by the jump in the relative energy contribution of oxygen for modes with higher frequencies than mode $F_{g}^{(5)}$.

Regarding the suitability to study oxygen related phenomena, Klein \textit{et al.} estimated a maximum concentration of oxygen vacancies of $3\cdot 10^{20}$ for In$_2$O$_3$,\citep{klein2024} translating into vacant oxygen-sites of approximately 1\%. In Wurtzite-type ZnO, such oxygen vacancy concentrations resulted in 1.6\% lower mode frequency for a mode associated with the vibration of the oxygen system.\citep{fukushima2015} Assuming a similar influence on the bixbyite In$_2$O$_3$ system and a line-fitting accuracy of 0.1~cm$^{-1}$, an analysable oxygen vacancy concentration down to 0.02\% could be resolved. Further research is necessary to investigate this potential usecase. The phonon mode $E_{g}^{(4)}$~cm$^{-1}$ appears to be most sensitive from a theoretical point of view, given the fact that only oxygen atoms contribute to this mode. Experimentally, however, the analysis of the $E_{g}^{(4)}$ mode is impracticable due to its extremely low intensity in real-world experiments. Similar arguments can be applied to mode $E_{g}^{(2)}$ and $A_{g}^{(4)}$, with only 1\% of the mode's energy contributed by indium lattice sites. Our recommendation for experimental probing of oxygen lattice site related effects, fall towards mode $F_{g}^{(14)}$. This mode has a relative small In contribution of 3\% and shows the largest occurring absolute shift in our isotopologues experiments, while having negligible overlap with other spectral features.

\section{Summary and conclusion}

We conducted a comprehensive theoretical and experimental investigation of phonon modes in cubic bixbyite In$_2$O$_3$ with varying oxygen isotopic compositions. Using DFPT simulations, we analysed crystals with pure $^{16}$O and $^{18}$O isotopes, as well as three intermediate mixtures, and confirmed these results with Raman spectroscopy measurements on a nearly pure $^{18}$O In$_2$O$_3$ thin film and a bulk $^{16}$O reference sample.
The DFPT simulations yielded several key findings. Except for a single pure oxygen vibration, all Raman-active phonon modes involve coupled indium-oxygen vibrations, with no indium-only vibrational modes observed. 
Most notably, two different plateaus in the relative isotopic mode shift are found for low-frequency and high-frequency phonon modes, respectively. This behaviour is explained by step-like increase in the energy contribution of oxygen to phonon modes with frequencies above 215~cm$^{-1}$.
Furthermore, the substitution of $^{16}$O with $^{18}$O leads to a linear decrease of a mode's Raman shift.
The experimental Raman spectra align closely with theoretical predictions, affirming the accuracy of the DFPT model, though with a slight underestimation of mode energies in the calculations. Our study establishes a clear linear relationship between the oxygen isotope ratio and phonon mode energies, highlighting the utility of Raman spectroscopy as a precise method for examining isotopic effects in cubic bixbyite crystals. Future research can leverage this approach to explore fundamental properties of In$_2$O$_3$ and to gain insights into crystal growth dynamics through controlled isotopic variation.

\section*{Conflicts of Interest}

There are no conflicts of interest to declare.

\section*{Acknowledgements}

We thank Tobias Schulz and Hans Tornatzky for a critical reading of the manuscript. This work was performed in the framework of GraFOx, a Leibniz-ScienceCampus partially funded by the Leibniz association. J.\ F.\ gratefully acknowledges the financial support by the Leibniz Association. The authors gratefully acknowledge the scientific support and HPC resources provided by the Erlangen National High Performance Computing Center (NHR$@$FAU) of the Friedrich-Alexander-Universit\"at Erlangen-N\"urnberg (FAU) under the NHR project b181dc. NHR funding is provided by federal and Bavarian state authorities. NHR$@$FAU hardware is partially funded by the German Research Foundation (DFG) 440719683.

\bibliography{In2O3_18O_JMC}
\bibliographystyle{rsc} 

\end{document}


\title{Supplemental Material: Unravelling the oxygen influence in cubic bixbyite In$_2$O$_3$ on Raman active phonon modes by isotope studies}

\author{Johannes Feldl}
\affiliation{Paul-Drude-Institut f\"ur Festk\"orperelektronik, Leibniz-Institut im Forschungsverbund Berlin e.\,V., Hausvogteiplatz 5--7, 10117 Berlin, Germany}
\author{Roland Gillen}
\affiliation{Chair of Experimental Physics, Friedrich-Alexander Universit\"at Erlangen-N\"urnberg, Staudtstr. 7, 91058 Erlangen, Germany}
\affiliation{College of Engineering, Swansea University, Swansea SA1 8EN, United Kingdom }
\author{Janina Maultzsch}
\affiliation{Chair of Experimental Physics, Friedrich-Alexander Universit\"at Erlangen-N\"urnberg, Staudtstr. 7, 91058 Erlangen, Germany}
\author{Alexandra Papadogianni}
\affiliation{Paul-Drude-Institut f\"ur Festk\"orperelektronik, Leibniz-Institut im Forschungsverbund Berlin e.\,V., Hausvogteiplatz 5--7, 10117 Berlin, Germany}
\author{Joe Kler}
\affiliation{Institute of Physical Chemistry, RWTH Aachen University, Aachen, Germany}
\author{Zbigniew Galazka}
\affiliation{Leibniz-Institut f\"ur Kristallz\"uchtung, Max-Born-Stra{\ss}e 2, 12489 Berlin, Germany}
\author{Oliver Bierwagen}
\email{bierwagen@pdi-berlin.de}
\affiliation{Paul-Drude-Institut f\"ur Festk\"orperelektronik, Leibniz-Institut im Forschungsverbund Berlin e.\,V., Hausvogteiplatz 5--7, 10117 Berlin, Germany}
\author{Manfred Ramsteiner}
\affiliation{Paul-Drude-Institut f\"ur Festk\"orperelektronik, Leibniz-Institut im Forschungsverbund Berlin e.\,V., Hausvogteiplatz 5--7, 10117 Berlin, Germany}

	\begin{abstract}
		{}
	\end{abstract}
	
	\maketitle
	\onecolumngrid

\begin{figure}[h]
\includegraphics*[width=1.0\textwidth]{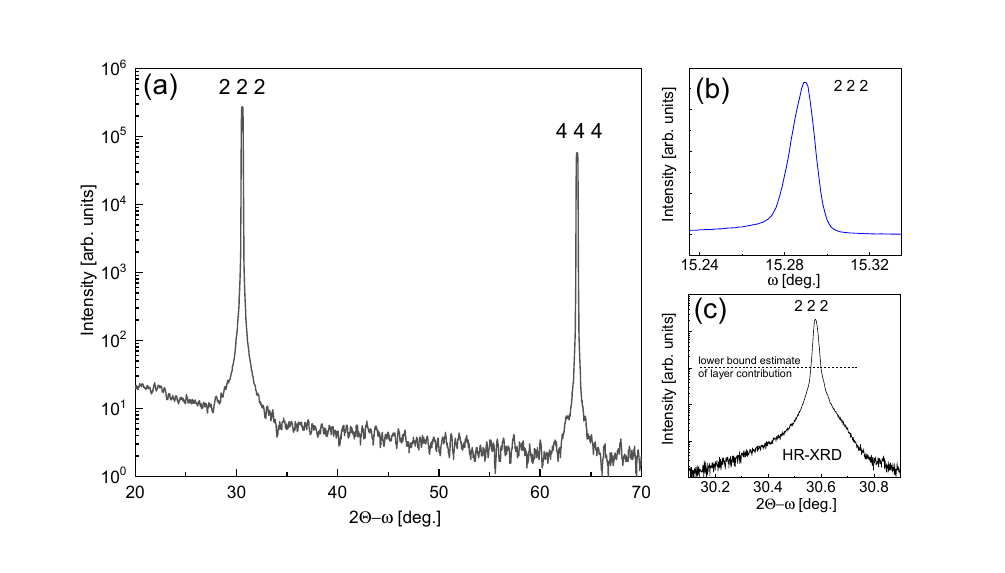}
\caption{Symmetric out-of-plane XRD analyis of the 500~nm-thick In$_2^{18}$O$_3$ layer on the In$_2$O$_3$ substrate. \textit{(a)} Wide range $2\Theta-\omega$-scan with logarithmic intensity scale, \textit{(b)} $\omega$-rocking curve of the 2~2~2 reflex with linear intensity scale, and \textit{(c)} high-resolution XRD $2\Theta-\omega$-scan of the 2~2~2 reflex with logarithmic intensity scale. The horizontal line shows a lower bound estimate for the intensity contribution of the layer. 
\label{sup_XRD}}
\end{figure}

Figure~\ref{sup_XRD} shows the symmetric out-of-plane XRD analysis of the 500~nm-thick homoepitaxial In$_2^{18}$O$_3$ layer. The wide-range $2\Theta-\omega$-scan (Fig.~\ref{sup_XRD}\textit{(a)}) shows the presence of two diffraction orders of the In$_2$O$_3$ (1 1 1) planes and the absence of other orientations or secondary phases, thus indicating epitaxial match of the film to the substrate. 
The narrow $\omega$-rocking curve (Fig.~\ref{sup_XRD}\textit{(b)}) with full-widths at half maximum of 0.014~deg is confirms a high crystal quality of substrate and film. The high-resolution XRD (Fig.~\ref{sup_XRD}\textit{(c)}) indicates the 2~2~2 reflex at $2\Theta=30.580$~deg, corresponding to a cubic lattice parameter of 1.0119~nm. The 2~2~2 reflex intensity measured on an In$_2^{18}$O$_3$ layer with same thickness grown in the same growth run on a co-loaded c-plane sapphire substrate was found to be $\approx5$\% that of the homoepitaxial layer. Thus, the reflex of the homoepitaxial layer may be dominated by the substrate contribution yet the layer should still contribute $\ge5$\%. At this intensity level (indicated by the  horizontal dashed line) the reflex looks symmetric with a widths of only 0.04~deg, suggesting the potential layer reflex to deviate no more than 0.01~deg from the substrate reflex (corresponding to a lattice mismatch or layer strain of $\le 0.035$\%).
\\

\begin{figure}[t]
\includegraphics*[width=0.75\textwidth]{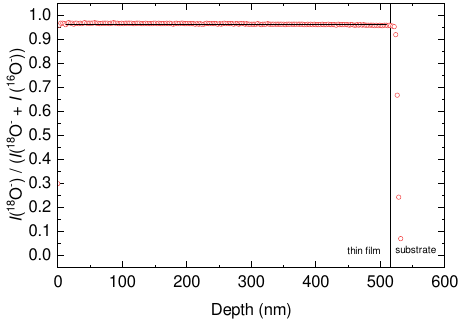}
\caption{ToF-SIMS depth profile of our $^{18}$O isotope MBE grown thin film on a $^{16}$O substrate, showing a 96.5\% $^{18}$O isotope fraction at the surface of the film, estimated by a linear fit (black line). The steep drop in $^{18}$O isotope fraction marks the film thickness at approximately 0.5~$\mu$m.} 
\label{sup_SIMS_18O}
\end{figure}

\begin{figure}[t]
\includegraphics*[width=0.65\textwidth]{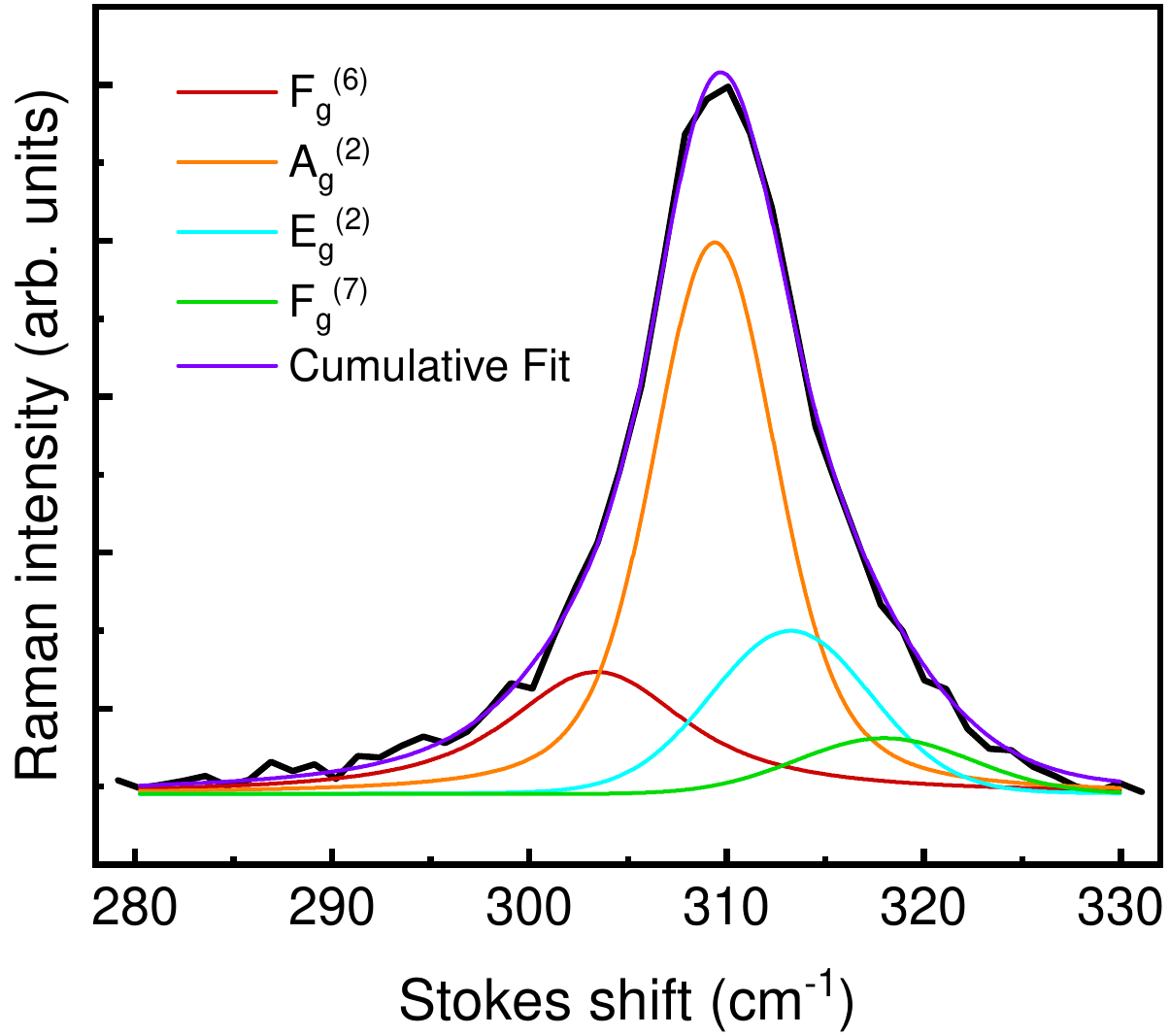}
\caption{Section of the $^{16}$O bulk sample Raman signal at 3.81~eV excitation energies, showing the superposition of $F_{g}^{(6)}$, $A_{g}^{(2)}$, $E_{g}^{(2)}$ and $F_{g}^{(7)}$ modes and their respective Voigt functions, derived by fitting procedures.} 
\label{sup_Eg2_fitting_details}
\end{figure}

For our experimental investigation of the influence of the oxygen isotope mass on the phonon mode frequency, it is crucial to have confirmation of the actual successfully incorporated oxygen isotopologue in the MBE grown thin film. For this purpose, we conducted an experiment independent of our Raman spectroscopic analysis tool by applying time-of-flight secondary ion mass spectrometry (ToF-SIMS). ToF-SIMS provides a quantitative depth profile of the isotope fraction of, in our case, oxygen ions. The data used for the analysis presented in the main article can be found in Figure~\ref{sup_SIMS_18O}. The sharp drop in $^{18}$O intensity is an indication of interface quality, with minimized diffusion despite elevated temperatures during growth. The drop occurs within a range of several tens of nanometers once the layer depth exceeds approximately 0.5~$\mu$m. An almost constant $^{18}$O isotope fraction of 96. 5\% was found throughout the film depth.
\\

To disentangle the Raman signal around 310~cm$^{-1}$, which arises from the superposition of four Raman modes ($F_{g}^{(6)}$, $A_{g}^{(2)}$, $E_{g}^{(2)}$ and $F_{g}^{(7)}$), we first analyzed the Raman spectrum of the $^{16}$O bulk sample using a 2.62~eV laser excitation. Under this excitation condition, Raman modes with $E_{g}$ symmetry exhibit significantly lower intensities due to their reduced scattering cross-section. Consequently, the measured spectrum primarily comprises contributions from the $F_{g}^{(6)}$, $A_{g}^{(2)}$ and $F_{g}^{(7)}$ modes.

Further discrimination of these modes was achieved through polarization optics, which enable symmetry-dependent intensity variations, allowing for an unambiguous assignment of the Raman peaks. The extracted peak positions are 303.4~cm$^{-1}$ for the $F_{g}^{(6)}$ mode, 309.4~cm$^{-1}$ for the $A_{g}^{(2)}$ and 318.0~cm$^{-1}$ for the $F_{g}^{(7)}$ mode. These positions were determined by fitting the spectrum with Voigt functions.

At an excitation photon energy of 3.81~eV, the analysis must also account for the $E_{g}^{(2)}$ mode, as its scattering cross-section is significantly enhanced and the experimental setup does not allow for symmetry induced suppression by polarization optics at this energy. To isolate the contribution of $E_{g}^{(2)}$, we performed spectral fitting while keeping the frequencies of the $F_{g}^{(6)}$, $A_{g}^{(2)}$, $E_{g}^{(2)}$ and $F_{g}^{(7)}$ modes fixed, with a correction applied for a setup specific offset. An additional Voigt function was introduced without constraints, allowing us to identify the $E_{g}^{(2)}$ mode at 313.3~cm$^{-1}$. The detailed fitting procedure is shown in Figure~\ref{sup_Eg2_fitting_details}.

\bibliography{In2O3_18O_supplemental.bib}